\documentclass[
reprint,
superscriptaddress,
amsmath,amssymb,
aps,
prl,
]{revtex4-2}

\usepackage{graphicx}% Include figure files
\usepackage{dcolumn}% Align table columns on decimal point
\usepackage{bm}% bold math
\usepackage{gensymb}
\begin{document}

\preprint{APS/123-QED}

\title{Anatomy of Thermally Interplayed Spin-Orbit Torque Driven Antiferromagnetic Switching}% Force line breaks with \\

\author{Wenlong Cai}
\thanks{These authors equally contribute to this paper}
\affiliation{Fert Beijing Institute, School of Integrated Circuit Science and Engineering, Beihang University, Beijing 100191, China}
\affiliation{National Key Laboratory of Spintronics, Hangzhou International Innovation Institute, Beihang University, Hangzhou 311115, China}
\author{Zanhong Chen}
\thanks{These authors equally contribute to this paper}
\author{Yuzhang Shi}
\thanks{These authors equally contribute to this paper}
\affiliation{Fert Beijing Institute, School of Integrated Circuit Science and Engineering, Beihang University, Beijing 100191, China}
\author{Daoqian Zhu}
\author{Guang Yang}
\affiliation{Fert Beijing Institute, School of Integrated Circuit Science and Engineering, Beihang University, Beijing 100191, China}
\affiliation{National Key Laboratory of Spintronics, Hangzhou International Innovation Institute, Beihang University, Hangzhou 311115, China}
\author{Ao Du}
\affiliation{Fert Beijing Institute, School of Integrated Circuit Science and Engineering, Beihang University, Beijing 100191, China}
\author{Shiyang Lu}
\affiliation{Fert Beijing Institute, School of Integrated Circuit Science and Engineering, Beihang University, Beijing 100191, China}
\author{Kaihua Cao}
\affiliation{Fert Beijing Institute, School of Integrated Circuit Science and Engineering, Beihang University, Beijing 100191, China}
\author{Hongxi Liu}
\affiliation{Truth Memory Tech. Corporation, China}
\author{Kewen Shi}
\email{shikewen@buaa.edu.cn}
\author{Weisheng Zhao}
\email{weisheng.zhao@buaa.edu.cn}
\affiliation{Fert Beijing Institute, School of Integrated Circuit Science and Engineering, Beihang University, Beijing 100191, China}
\affiliation{National Key Laboratory of Spintronics, Hangzhou International Innovation Institute, Beihang University, Hangzhou 311115, China}

\date{\today}% It is always \today, today,

\begin{abstract}
Current-induced antiferromagnetic (AFM) switching remains critical in spintronics, yet the interplay between thermal effects and spin torques still lacks clear clarification. Here we experimentally investigate the thermally interplayed spin-orbit torque induced AFM switching in magnetic tunnel junctions via pulse-width dependent reversal and time-resolved measurements. By introducing the Langevin random field into the AFM precession equation, we establish a novel AFM switching model that anatomically explains the experimental observations. Our findings elucidate the current-induced AFM switching mechanism and offer significant promise for advancements in spintronics.
\end{abstract}

\maketitle

Antiferromagnets (AFMs) are drawing great attention in the design of fast and scalable spintronic devices due to their advantages over conventional ferromagnets (FMs), such as the absence of stray fields, high robustness against external fields, and terahertz spin dynamics~\cite{Song2023PhysRevLett,chen2023nature,qin2023Nature,cai2023SCPMA,Kamil2018SA}. To achieve information regulation in AFMs,  it is essential to effectively control the orientation of their N\'{e}el vector through electrical manipulation~\cite{RevModPhys.90.015005,arpaci2021NC,Amiri2024AM}. Recently, spin-orbit torque (SOT) switching of AFMs has been extensively investigated~\cite{peng2020NE,Liu2022PRL,Liu2023NM}. The AFMs with locally broken inversion symmetry in their sublattices, such as $\rm{Mn}_2$Au and CuMnAs, were first theoretically predicted and experimentally demonstrated to be switchable by staggered spin torques~\cite{vzelezny2014PRL,science2016,grzybowski2017PRL,bodnar2018NC}. Subsequently, in AFM/heavy metal bilayers as shown in Fig. 1(a), it was found that the N\'{e}el vector (\textit{\textbf{L}}) of AFMs can be efficiently switched by SOT generated from neighboring heavy metal~\cite{chen2018PRL,baldrati2020PRL,takeuchi2021PRL,wang2022NE,lai2019NM,kang2021NC}. To clearly monitor AFM orders, a three-terminal exchange bias magnetic tunnel junction (EB-MTJ) using FM/AFM as the free layer was proposed to effectively read out the AFM state after current writing~\cite{fang2022AFM,zhu2021IEDM,du2023NE,du2024AEM}. It was theoretically predicted that the \textit{\textbf{L}}  could be driven to an oscillatory state by SOT~\cite{gomonay2010PRB,khymyn2017SR} as shown in Fig. 1(b) and then switched to the other side by the interfacial coupling between FM and AFM ~\cite{du2023NE}. Apart from the SOT effect, other factors during the application of the writing current, such as thermal effects, also play important roles in determining the switching of AFMs~\cite{meinert2018PRA,Baldrati2019PRL,Xu2022PRL}. However, the mechanism of current induced AFM switching, involving these multiple factors, remains poorly elucidated. Further research on its switching mechanism is highly required.

\begin{figure}[t]
\includegraphics[width=\columnwidth]{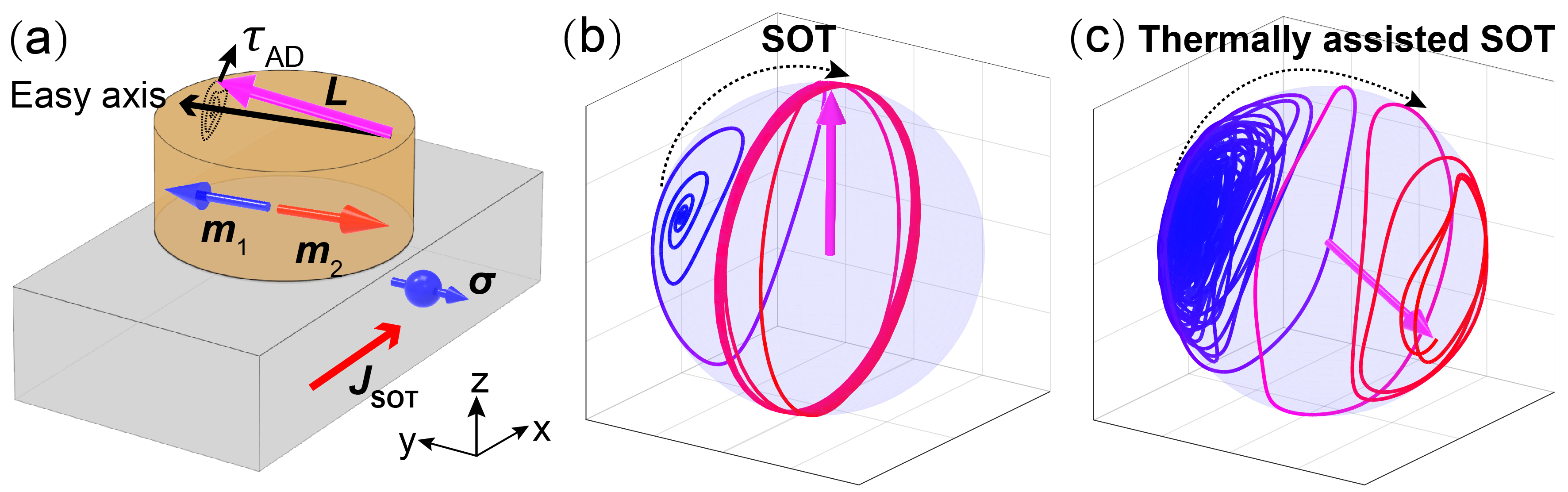}% Here is how to import EPS art
\caption{\label{Fig. 1} (a) Anti-damping of SOT ($\rm{\bm{\tau}}_{AD}$) induced rotation of the \textit{\textbf{L}} around the easy axis direction. (b) Precession trajectory of the \textit{\textbf{L}} due to SOT, ending with a full-angle precession in the plane near y = 0. (c) Stochastic precession trajectory motivated by thermal fluctuations and smaller SOT, undergoing sudden changes in precession angle due to thermal perturbations, and probabilistically switching into the other side.}
\vspace{-0.6cm} 
\end{figure}

In this Letter, we theoretically and experimentally clarify the mechanism of thermally interplayed SOT-induced AFM switching, as illustrated in Fig. 1(c). The asynchronous switching of AFMs and FMs is first investigated in EB-MTJs, enabling the accurate characterization of AFM switching. Meanwhile, based on the framework of the Landau-Lifshitz-Gilbert (LLG) equation with a Langevin random field for finite temperature, we develop an AFM switching model including current-induced temperature increase, SOT effect, and thermal activation. Moreover, through elaborate time-resolved measurements, we quantify the temperature rise during the switching process. Finally, the threshold current density for AFM switching with varying pulse widths and accurate device temperatures is explored, closely aligning with the predictions from the derived macroscopic model. Our findings clearly elucidate the mechanism of thermally interplayed SOT-induced AFM switching.

The multilayer stacks are deposited on thermally oxidized silicon substrates by DC/RF magnetron sputtering. The core MTJ structure consists of IrMn/CoFeB/MgO/CoFeB/Ru/CoFe/IrMn, with the CoFeB adjacent to IrMn serving as the free switching layer. Additionally, 8-nm-thick Pt and Cu buffer layers are deposited underneath the IrMn/CoFeB layer to serve as different spin sources for comparison, denoted as Pt/IrMn-MTJ and Cu/IrMn-MTJ, respectively. The spin Hall angle ($\xi_\text{DL}$) of Pt is estimated as 0.15(1) by harmonic measurements~\cite{supplementary,wen2017PRB}. The films are annealed at 300 °C for 1 hour under a magnetic field of 1 T along the y-direction. The tunneling magnetoresistance (TMR) effect is employed to read out the antiferromagnetism of IrMn layers. Subsequently, the nanopillars with a diameter of 700 nm are fabricated with a TMR ratio of ~100\% as shown in Fig. 2 (a). It is apparent that the exchange bias field ($\textit{H}_{\rm{EB}}$) is larger than the coercive field ($\textit{H}_{\rm{C}}$), enabling the storage functionality of EB-MTJs~\cite{du2023NE}. Afterward, the current-induced magnetization switching of Pt/IrMn-MTJ is implemented as shown in the insert of Fig. 2(b). Explicit switching occurs with an adequate current density ($\sim$ -120 MA cm$^{-2}$). However, this only represents the switching of the FM layer, while the threshold current for AFM switching remains unknown. 

Therefore, measurements with a combination of magnetic field and current are performed to explore the AFM switching. We first choose a series of writing current lower than that of FM switching, as shown in Fig. 2(b), and define the corresponding states of the free layer as 1 to 5, respectively. Then, we measure its \textit{R-H} curves with applying the writing currents as shown in Fig. 2(c). These curves indicate that $\textit{H}_{\rm{EB}}$ gradually decreases with increasing current until it vanishes. Figure 2(d) visually depicts the possible states of AFM and FM under different writing currents. The yellow arrow pairs represent the nonuniform stochastic precession of AFM orders. However, when a writing current is applied, the variation of $\textit{H}_{\rm{EB}}$ arises not only from the AFM precession but also from the coupling energy changes due to temperature increases, causing $\textit{H}_{\rm{EB}}$ to be an unreliable indicator of the AFM state in this experiment. Therefore, we apply an $\textit{H}_{\rm{y}}$ of -100 mT during the current writing to switch the FM magnetization into -y direction. Upon cessation of the writing current, the precessing AFM eventually stabilizes, reestablishing an EB field parallel to the FM magnetization, as shown in Fig. 2(e). The states 1-5 are transformed as states 1'-5', respectively. The corresponding \textit{R-H} curves shown in Fig. 2(f) display a gradual transition of $\textit{H}_{\rm{EB}}$, which can accurately represent the successive switching of AFM grains of different sizes. These results indicate that in Pt/IrMn-MTJs, the threshold current for FM switching is larger than that for AFM switching which exhibits gradual variation, necessitating the definition of certain states to study the characteristics of the current-induced AFM reversal.

\begin{figure}[t]
\includegraphics[width=\columnwidth]{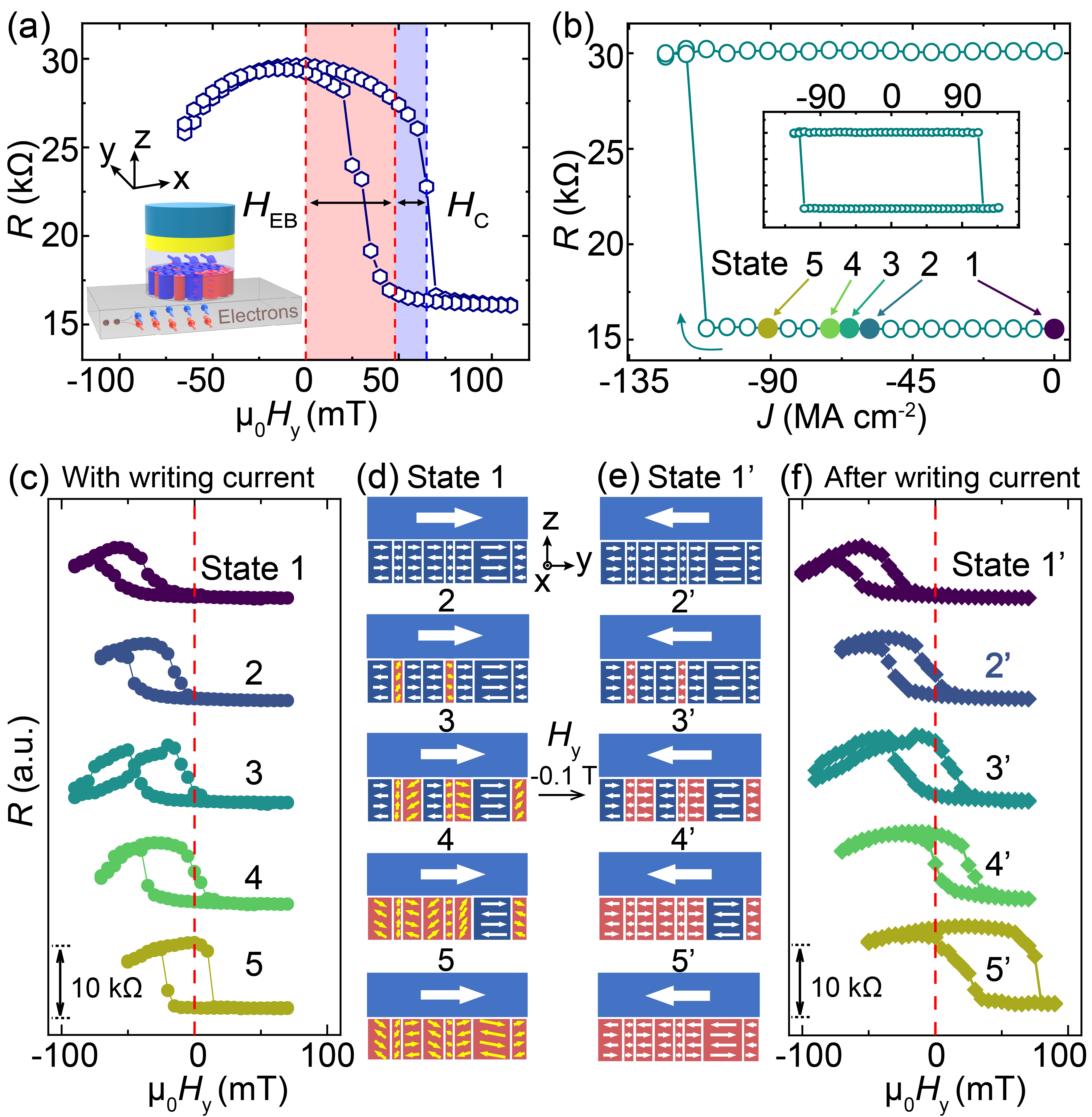}% Here is how to import EPS
\vspace{-0.2cm} 
\caption{\label{Fig. 2} (a) TMR of an EB-MTJ versus the y-direction magnetic field ($\textit{H}_{\rm{y}}$). Insert is the schematic of the devices. (b) Electrical switching of the device with states 1-5 corresponding to different writing currents. Insert shows the full curve of electrical switching. (c) \textit{R-H} loops of states 1-5 when applying the writing currents and (d) the corresponding states of FM and AFM. (e) Reformed states after applying $\mu_0\textit{H}_{\rm{y}}$= -0.1 T to align the FM magnetization into -y direction and then ceasing the writing current. (f) \textit{R-H} loops of related reformed states, i.e. 1'-5' after current writing.}
\vspace{-0.6cm} 
\end{figure}

Consequently, to estimate the threshold current for AFM switching, we adopt the aforementioned AFM writing methodology involving magnetic fields and currents. By increasing current, $\textit{H}_{\rm{EB}}$ progressively transitions from negative to positive, as illustrated in Fig. 3(a). During this process, the resistance of the MTJ device exhibits two abrupt changes when in a quiescent state (devoid of external field and current). The first resistance change happens when $\textit{H}_{\rm{EB}}$ = -$\textit{H}_{\rm{C}}$ by applying a writing current under the $\textit{H}_{\rm{y}}$ of -100 mT, corresponding to the state 2' in Fig. 2(f). The middle panels of Figs. 3(b) and (c) display this switching in Pt/IrMn-MTJs and Cu/IrMn-MTJs, respectively. If we continue to increase the writing current, we can observe the second resistance change when $\textit{H}_{\rm{EB}}$ = $\textit{H}_{\rm{C}}$ by applying another positive $\textit{H}_{\rm{y}}$ (100 mT) after the current writing and then detecting the final resistance of the devices, corresponding to the state 4' in Fig. 2(f). The switching threshold current density can be efficiently characterized as shown in the bottom panels of Figs. 3(b) and (c).

Here, we define the current density for reversing the AFM magnetization in the majority of grains, \textit{J}($\textit{H}_{\rm{EB}}$ = 0), as the threshold for AFM switching($\textit{J}_\text{C}$), expressed as [\textit{J}($\textit{H}_{\rm{EB}}$ = $\textit{H}_{\rm{C}}$)+\textit{J}($\textit{H}_{\rm{EB}}$ = $\textit{-H}_{\rm{C}}$)]/2, corresponding to states 3', 4', and 2' in Fig. 2(f). By varying the pulse width ($\tau_\text{PW}$) of the writing current, we measure the threshold current of both FM and AFM in Pt/IrMn-MTJs and Cu/IrMn-MTJs, as shown in Figs. 3(d) and 3(e), respectively. Regardless of pulse width, the AFM switching current of Pt/IrMn-MTJs is consistently lower than that of FM switching. For Cu/IrMn-MTJs, the AFM switching current is slightly lower than that of FM when $\tau_\text{PW} >$ 100 ns and becomes almost the same when $\tau_\text{PW}$ is further deceased. For $\tau_\text{PW} <$ 100 ns, when applying a current slightly below the threshold, the FM magnetization should be already reversed, while still keeping an unchanged final state due to the pinning effect from the stable AFM. In other words, the threshold current at  $\tau_\text{PW} <$ 100 ns for independent FM switching is lower than that for AFM switching in Cu/IrMn-MTJs. Additionally, for AFM switching, a linear relationship between the threshold current density and the natural logarithm of $\tau_\text{PW}$ is observed when $\tau_\text{PW}$ exceeds 1 \textmu s. Conversely, below 1 \textmu s, the threshold current density increases significantly, suggesting a complex mechanism governing current-induced AFM switching. 

\begin{figure}[t]
\includegraphics[width=\columnwidth]{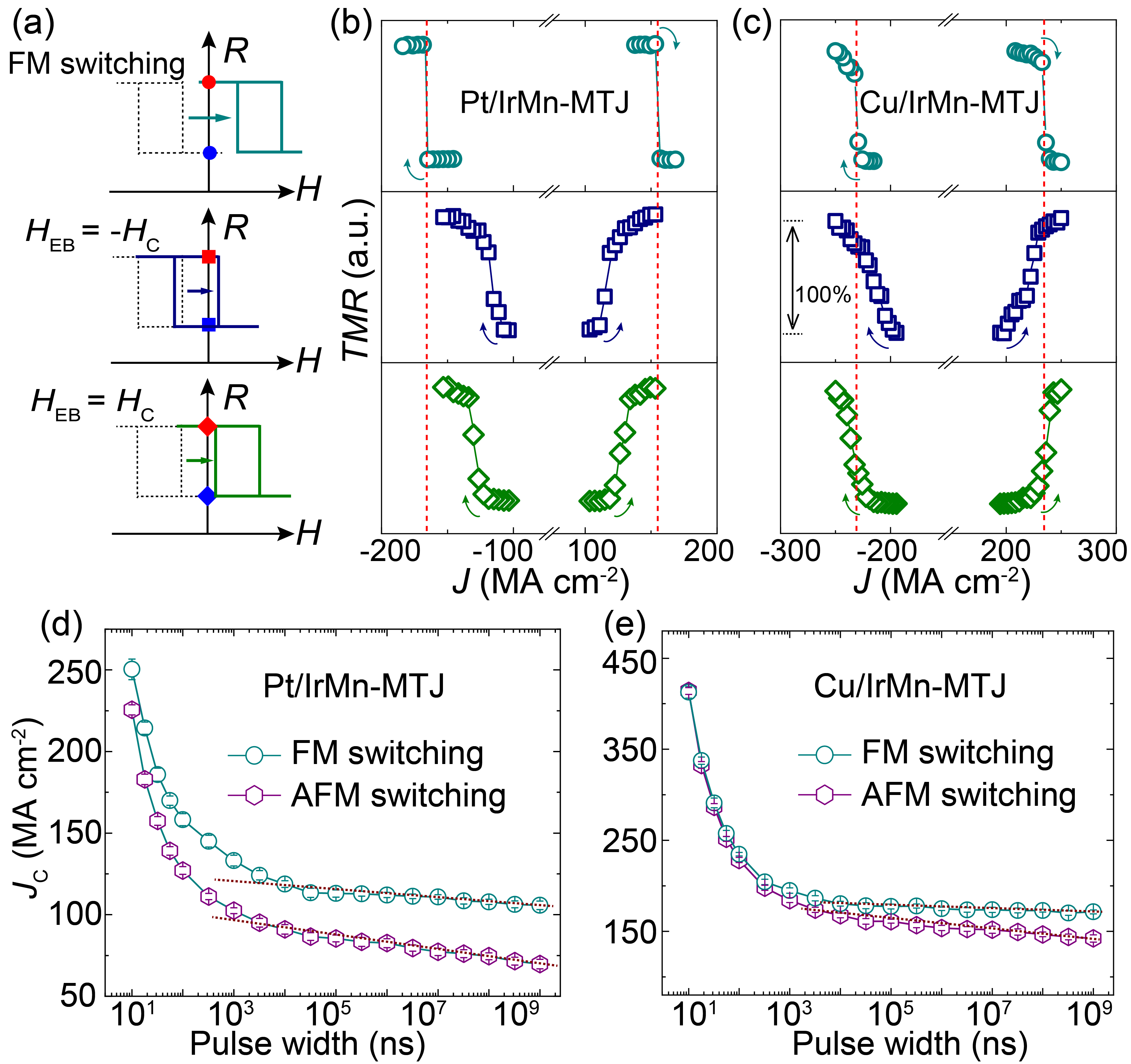}% Here is how to import EPS art
\vspace{-0.2cm} 
\caption{\label{Fig. 3} (a) Diagrams of FM switching and two states of exchange bias reversal, $\textit{H}_{\rm{EB}}$ = $\textit{H}_{\rm{C}}$ and $\textit{H}_{\rm{EB}}$ = -$\textit{H}_{\rm{C}}$, respectively. Corresponding FM and AFM asynchronous switching curves in (b) a Pt/IrMn-MTJ and (c) a Cu/IrMn-MTJ. The pulse width here is 100 ns. Threshold current density ($\textit{J}_\text{C}$) for FM and averaged AFM switching versus pulse widths ranging from 10 ns to 1 s in (d) the Pt/IrMn-MTJ and (e) the Cu/IrMn-MTJ.}
\vspace{-0.5cm} 
\end{figure}

To get insight into the mechanism of AFM switching, we build a thermally interplayed SOT-induced AFM switching model. Starting from the LLG equation for AFM sublattice ($\textbf{\textit{M}}_1$, $\textbf{\textit{M}}_2$), the precession equation for the antiferromagnetic N\'{e}el vector (\textbf{\textit{L}} = $\textbf{\textit{M}}_1$ - $\textbf{\textit{M}}_2$) can be derived in generalized coordinates $(L_{\mathrm{x}}, L_{\mathrm{y}}, L_{\mathrm{z}})$~\cite{gomonay2010PRB,supplementary}:
\begin{eqnarray}
\hspace{-0.6cm} 
\ddot{L}_{\mathrm{x}} + 2\gamma_{\mathrm{AFM}} \dot{L}_{\mathrm{x}} + \omega^2 L_{\mathrm{x}} - \gamma \mu_0 H_{\mathrm{E}} \sigma_{\text{SOT}} J L_{\mathrm{z}} &=& 0,
\end{eqnarray}
\begin{eqnarray}
\hspace{-0.6cm} 
\ddot{L}_{\mathrm{z}} + 2\gamma_{\mathrm{AFM}} \dot{L}_{\mathrm{z}} + \omega^2 L_{\mathrm{z}} + \gamma \mu_0 H_{\mathrm{E}} \sigma_{\text{SOT}} J L_{\mathrm{x}} &=& 0,
\end{eqnarray}
where $\gamma_{\text{AFM}}$ is defined as $\gamma\mu_0 \textit{H}_\text{E}\alpha_\text{G}/2$, $\textit{H}_\text{E}$ is the exchange field, $\alpha_\text{G}$ is the Gilbert damping, $\omega$ is expressed as $2 \gamma \mu_0 \sqrt{H_{\text{an}}\textit{H}_\mathrm{E}}$,  $H_{\text{an}}$ is the magnetic anisotropy field, $\sigma_\text{SOT} = \xi_\text{DL}\hbar\gamma/(2\textit{M}_0 t_\text{AFM}e)$, $t_\text{AFM}$ is the thickness of AFMs, and $\textit{M}_0$ is the saturation magnetization of the sublattice. Equation (1) can be solved as $L_{\mathrm{x, z}}=\delta e^{-\gamma_{\text{AFM}} t + \frac{  \sigma_\text{SOT} J t}{4} \sqrt{{H_\text{E}}/{H_{\text{an}}}}} e^{i(\omega t + \phi_{\mathrm{x, z})}}$~\cite{supplementary}, where $\delta$ is the slight deviation of the N\'{e}el vector's initial position from the precession axis, $\phi_\text{x}$ and $\phi_\text{z}$ are the phase angles of $\textit{L}_\text{x}$ and $\textit{L}_\text{z}$, respectively, satisfying $\phi_\text{z}-\phi_\text{x} = \pi/2$. $L_\mathrm{y}$ can be written as $2M_0 - \left({L_\mathrm{x}^2 + L_\mathrm{z}^2}\right)/{4M_0}$. By differentiating the expression of $\textit{\textbf{L}}$ with respect to time, we can obtain the following equation~\cite{supplementary}:
\begin{eqnarray}
\hspace{-0.5cm} 
\dot{\bm{L}} &=& -\bm{L} \times \bm{\omega} - A_{\mathrm{\alpha}} \bm{L} \times (\bm{L} \times \bm{\omega}) + A_J \bm{L} \times (\bm{L} \times \bm{\sigma}),
\end{eqnarray} 
where $\bm{\omega}$ represents the angular velocity of the N\'{e}el vector with the generalized coordinates (0, $\omega$, 0), $A_{\mathrm{\alpha}} \equiv \frac{\alpha_\text{G}}{8 M_0} \sqrt{\frac{H_\text{E}}{H_{\text{an}}}}$ is the coefficient indicating the damping effect, and $A_J \equiv \frac{\sigma_\text{SOT} J}{8 M_0} \sqrt{\frac{H_\text{E}}{H_{\text{an}}}}$ represents the SOT effect on the AFM precession. To account for the influence of thermal fluctuations during the N\'{e}el vector switching, we introduce a Langevin random field $\textit{\textbf{H}}_\text{Lr}$ to the effective magnetic field ($\bm{\omega}/\gamma$) related to the system temperature as $\textit{H}_\text{Lr,i} = \sqrt{2{A_{\alpha} k_\text{B} T}/{\gamma}} G_{\text{ran,i}}(\textit{t})$ (i = x, y, z), where $G_{\text{ran}}(\textit{t})$ is a Gaussian random function with a mean of $<G_\text{ran}(\textit{t})> = 0$ and variance of  $<G^2_\text{ran}(\textit{t})> = 1$~\cite{brown1963PR,grinstein2003PRL,koch2004PRL}. 

Furthermore, given that $\bm{\sigma}$ is aligned with $\bm{\omega}$, the equation describing the precession of the N\'{e}el vector, including the thermal term $\textit{\textbf{H}}_\text{Lr}$, can be written as: 
\begin{eqnarray}
\dot{\textbf{\textit{L}}} &=& -\textbf{\textit{L}} \times (\bm{\omega} +\gamma\mu_0 \textbf{\textit{H}}_{\text{Lr}}) - \tilde{A}_{\alpha} \textbf{\textit{L}} \times (\textbf{\textit{L}} \times \bm{\omega}),
\end{eqnarray} 
where $\tilde{A}_{\alpha} \equiv (\frac{\alpha_\text{G}}{8 
 M_0}-\frac{\sigma_\text{SOT} J}{8 M_0 \omega}) \sqrt{\frac{H_\text{E}}{H_{\text{an}}}}$ represents the effective damping coefficient including the SOT term. Here, $\textbf{\textit{H}}_{\text{Lr}}$ can be written as $\textit{H}_\text{Lr,i} = \sqrt{2{\tilde{A}_{\alpha} k_\text{B} \tilde{T}}/{\gamma}} G_{\text{ran, i}}(\textit{t})$ (i=x, y, z), describing the Langevin random field equivalent with a damping coefficient $\tilde{A}_{\alpha}$ at the effective temperature $\tilde{T}$ with the equation $\tilde{\alpha}\tilde{T}=\alpha T$. Therefore, $\tilde{T}$ has the expression as $T/[1-J/(\alpha_\text{G}\omega/\sigma_\text{SOT})]$. More importantly, the thermally assisted SOT-induced AFM switching model can be described by the lifetime of the activated AFM N\'{e}el vector using the equation:
 \begin{eqnarray}
\tau = \tau_0 e^{\frac{E_\text{an}}{k_\text{B} \tilde{T}}} = \tau_0 e^{\frac{E_\text{an}}{k_\text{B} T}\left(1 - \frac{J}{J_\text{C0}}\right)},
\end{eqnarray}
where $\tau_0$ is the inverse of the attempt frequency of AFMs, $E_\text{an}$ is the magnetic anisotropy energy of AFMs, and $J_\text{C0} \equiv \alpha_\text{G}\omega_z/\sigma_\text{SOT}$ shows the same formula with the threshold current keeping up a stable rotation of N\'{e}el vector around $\bm{\sigma}$~\cite{gomonay2010PRB}.

According to the model derived above, the linear trends of the thresholds above 1 \textmu s in Fig. 3(d) align well with the formula. However, below 1 \textmu s, the threshold current deviates from the model, indicating additional effects from the writing current. To accurately describe the multiple influences during AFM switching, such as SOT, temperature rise, and thermally activated probabilistic switching, Equation (4) is derived as follows:
\begin{eqnarray}
\tau &=& \tau_0 e^{\frac{K_\text{an}V}{k_\text{B} T}\left(1 - \frac{T}{T_\text{N}}\right)\left(1 - \frac{J}{J_\text{C0}}\right)},
\end{eqnarray} 
where $K_\text{an}$ is the magnetic anistropy energy density of AFMs at 0 K satisfying $E_\text{an}(T) = K_\text{an}V(1-{T}/{T_\text{N}})$~\cite{vallejo2010APL} and $T_\text{N}$ is the N\'{e}el temperature of AFMs. The AFM switching model described by Equation (5) can be phenomenologically represented by the energy barrier schematic in Fig. 4(a).

 Considering the non-uniform shape of the bottom electrode, significant deviations may occur when used as a temperature sensor~\cite{supplementary}. To mitigate this issue, we use the MTJ resistance to characterize the real-time device temperature. We characterize the dependence of the MTJ resistance on the temperature~\cite{supplementary} and writing current separately to establish the relationship between the writing current and the device temperature as shown in the insert of Fig. 4(b). The symmetry of resistance variations indicates that changes are primarily due to temperature rises. Then, we perform time-resolved measurements with an oscilloscope to detect real-time resistance by dividing the background signals~\cite{supplementary}, as shown in Fig. 4(b), reflecting the real-time temperature changes. It can be observed that the temperature rises rapidly due to the Joule heating of the writing current within microsecond time scale \cite{grimaldi2020NN,cai2021EDL} and gradually reaches its saturation. 

Based on the real-time estimation, we can accurately determine the corresponding device temperatures. Particularly for $\tau_\text{PW}<$1 \textmu s, the temperature does not reach saturation. By integrating the switching probabilities~\cite{supplementary}, we obtain the effective temperatures ($\textit{T}_\text{switching}$) during the current induced AFM switching for different $\tau_\text{PW}$, as shown in Fig. 4(c). Subsequently, we characterize the blocking temperature ($\textit{T}_\text{b}$) over different testing durations, as indicated by the maroon circles~\cite{supplementary}. Notably, characterizing $\textit{T}_\text{b}$ involves using $\textit{H}_\text{EB}=0$ as the criterion for determination, the same as for the aforementioned current-induced AFM switching. The fitting curves, which follow the thermal equation~\cite{vallejo2010APL}, demonstrate that $\textit{T}_\text{switching}$ in the Cu/IrMn-MTJ closely matches their $\textit{T}_\text{b}$ at different time scales. This suggests that AFM switching in these devices is primarily driven by Joule heating, with negligible SOT influence due to the weak spin-orbit coupling effect of Cu. In contrast, in the Pt/IrMn-MTJ device, $\textit{T}_\text{switching}$ is noticeably lower than $\textit{T}_\text{b}$. Furthermore, as $\tau_\text{PW}$ decreases, the switching current significantly increases, while the Joule heating effect gradually diminishes, indicating a substantial increase in the contribution of SOT. 

Then, we incorporate the temperature change of the Pt/IrMn-MTJ into the thermally assisted SOT switching model derived above. The normalized pulse width corrected with a temperature term is used as the horizontal axis, while the threshold SOT current density for AFM switching is plotted on the vertical axis as shown in Fig. 4(d). The clear linear dependence, consistent with Equation (5), validates the correctness of the thermally interplayed SOT-induced AFM switching model. Moreover, the intrinsic threshold SOT current density for the AFM switching can be determined as 338 MA cm$^{-2}$ from the fitting data of the Pt/IrMn-MTJ, which is approximately an order of magnitude larger than that for FM switching~\cite{miron2011nature}.

\begin{figure}
\includegraphics[width=\columnwidth]{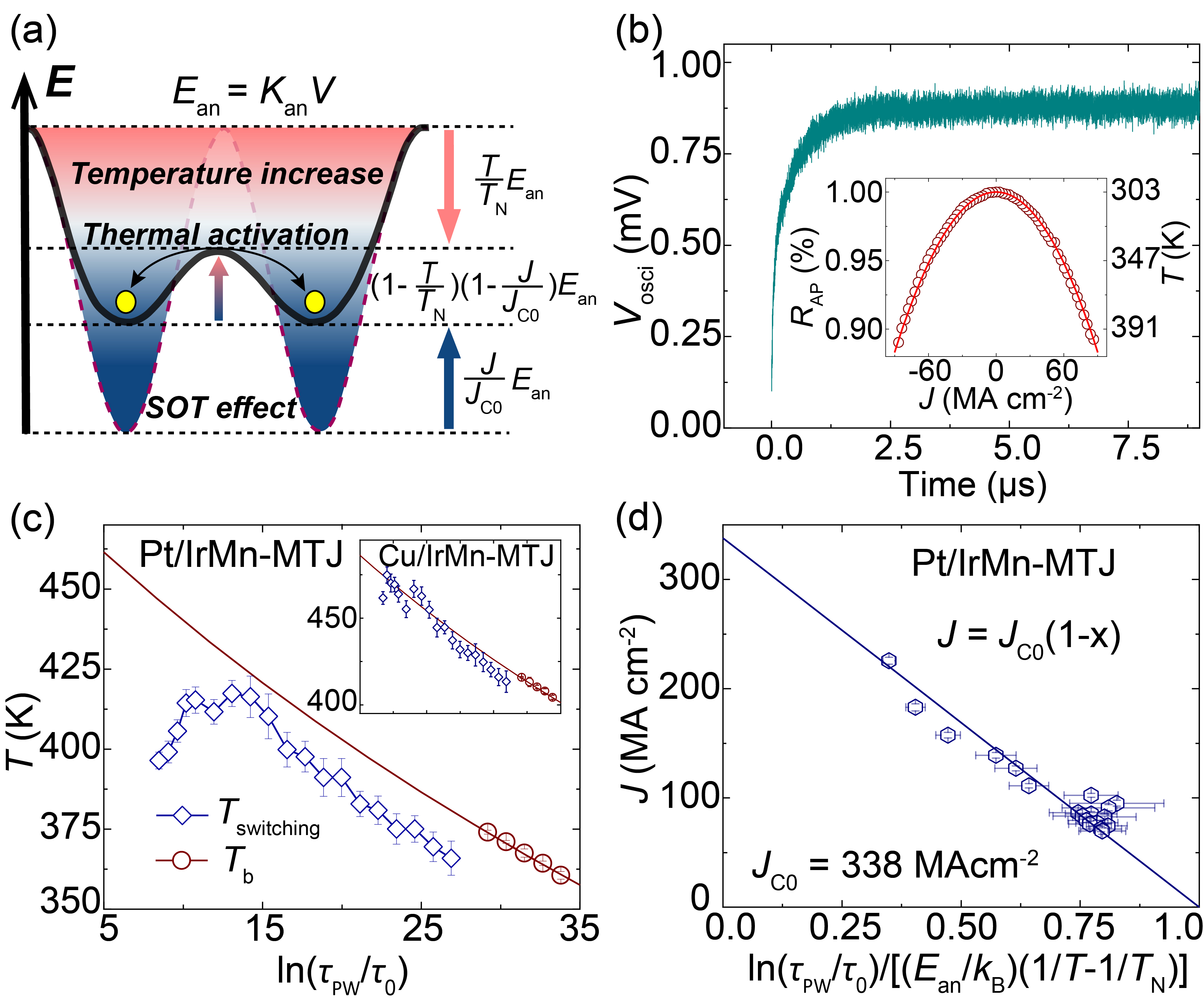}% Here is how to import EPS
\vspace{-0.3cm} 
\caption{\label{Fig. 4} (a) Schematic of AFM switching across potential valleys over energy barriers, induced by current-driven temperature increases, the SOT effect, and thermal activation. (b) Time-resolved voltage variations resulting from changes in MTJ resistance due to temperature increases. The inset displays the percentage change in antiparallel MTJ resistance ($\textit{R}_\text{AP}$) and the corresponding temperature versus writing current. (c) Blocking and estimated temperatures during AFM switching in the Pt/IrMn-MTJ, with inset details for the Cu/IrMn-MTJ. (d) SOT threshold current density versus corrected time, showing a linear dependence.}
\vspace{-0.6cm} 
\end{figure}

In conclusion, we derive and experimentally validate a physical model for thermally interplayed SOT-induced switching of AFMs. Theoretically, by introducing SOT and a Langevin random field into the LLG equation framework, we build a macroscopic model for thermally assisted SOT-induced AFM switching. Experimentally, we initially observe asynchronous switching between FMs and AFMs in EB-MTJs and undertake a comprehensive study on pulse width-dependent switching. Furthermore, by quantifying real-time temperature increases, we perform a detailed analysis that isolates Joule heating's contribution to the thermally interplayed SOT-induced AFM switching, aligning our findings with the developed physical model. This work elucidates the mechanism behind thermally interplayed SOT-induced AFM switching, paving the way for further research and advancements in antiferromagnetic spintronics.

This work is supported by the National Key Research and Development Program of China (Grant No. 2022YFB4400200), the National Natural Science Foundation of China (Grant Nos. 62271026, T2394473, T2394474, and T2394470), National Postdoctoral Program for Innovative Talents, and the China Postdoctoral Science Foundation (Grant No. 2023M740177). The electrical detection setups are supported in part by Truth Instruments Co., Ltd.

\nocite{*}

\bibliography{references}% Produces the bibliography via BibTeX.

\end{document}